\documentstyle[12pt]{article}
\advance\textheight by \topskip
\begin{document}
\begin{center}
\begin{flushright}
CERN-TH/95-290\\
hep-th/9511056
\end{flushright}
\vspace{1cm}
{\large\bf  T-DUALITY AND NON-LOCAL REALIZATIONS OF\\
                SUPERSYMMETRY IN STRING THEORY
\footnote{{\rm Talk presented at the Workshop on Strings, Gravity
  and Related Topics, Trieste, 29-30 June 1995}}
}\\
\vskip 1.5 cm
{\bf S. F. Hassan}\footnote{ E-mail address:
{\tt fawad@surya1.cern.ch}}
\vskip 0.1cm
{\it Theory Division, CERN \\ CH-1211 Geneva 23, Switzerland}\\
\end{center}
\vskip 2cm
\centerline{\bf ABSTRACT}
\begin{quotation}
\noindent
We study non-local realizations of extended worldsheet
supersymmetries and the associated space-time supersymmetries
which arise under a T-duality transformation. These non-local
effects appear when the supersymmetries do not commute with the
isometry with respect to which T-duality is performed.
\end{quotation}
\vfil
\begin{flushleft}
CERN-TH/95-290\\
November, 1995
\end{flushleft}
\newpage
%Some definitions
\def\odd{$O(d,d)$}
\def\oddh{$O(d,d+16)$}
\def\be{\begin{equation}}
\def\ee{\end{equation}}
\def\beq{\begin{eqnarray}}
\def\eeq{\end{eqnarray}}
\def\wt{\widetilde}
\def\del{\partial}
\def\s{$\sigma$}
\renewcommand{\Large}{\large}
\section{Introduction}
In string theory, a T-duality transformation is usually performed with
respect to a space-time coordinate (say $\theta$), provided the massless
background fields are invariant under translations in the
$\theta$-direction. The dual theories are different space-time
manifestations of the same superconformal field theory. To study the
full theory, one has to go beyond the massless background fields and
find how other objects and vertex operators in the theory map under
duality. It so happens that when an object is not invariant under a
$\theta$-translation, then in the dual theory, it is always realized
non-locally\cite{K}. This effect can be easily studied when
T-duality is implemented by a canonical transformation in the
worldsheet theory\cite{GRV}. Of particular interest is the behaviour
of extended worldsheet supersymmetries and their associated target
space supersymmetries under a T-duality transformation. When the
supersymmetry charges are invariant under translations in the
coordinate $\theta$, then a duality with respect to $\theta$ does not
give rise to non-local effects and the supersymmetries are
preserved\cite{H1,KIR,BOK}. In \cite{H2} (see also\cite{BS}), we
studied the general situation where the supercharges can depend on
$\theta$ and addressed the issue of non-local realizations of
supersymmetry. This talk contains a summary of the results appearing
in this paper and is organised a follows: First, we formulate
T-duality as a canonical transformation in an $N=1$ supersymmetric
non-linear $\sigma$-model. Then we consider theories with extended
supersymmetry on the worldsheet and obtain the non-local objects which
replace the $\theta$-dependent complex structures in the dual theory.
Using these results, we investigate the effects of this non-locality
on the associated target space supersymmetry.
\section{T-Duality as a Canonical Transformation in Supersymmetric
  Theories}
Let us consider massless bosonic background fields $G_{MN}, B_{MN}$
($M,N=1,..., D$) and $\Phi$, which do not depend on one of the target
space coordinates, denoted by $X^1=\theta$, but may have a
dependence on the remaining coordinates $X^{i+1}=x^i; i=1,...,D-1$.
Under a T-duality transformation with respect to $\theta$, the
background fields $G_{MN}$ and $B_{MN}$ transform as
\beq
&\wt G_{\theta \theta} = G^{-1}_{\theta\theta},\qquad
(\wt G\pm\wt B)_{\theta i}=\mp G^{-1}_{\theta\theta}(G\pm B)_{\theta i},&
\nonumber\\
\label{dual}
&(\wt G+\wt B)_{ij}=(G+B)_{ij}-
G^{-1}_{\theta\theta}(G-B)_{\theta i}(G+B)_{\theta j}.&
\eeq
To write down the transformation of the torsionful connections
$\Omega^{\pm K}_{MN}=\Gamma^K_{MN} \pm {1\over 2} G^{KL}H_{LMN}$
under duality, we introduce two $D\times D$ matrices $Q_\pm$ given by
\cite{H1}:
\be
\label{Qs}
Q_\pm =\left(
\begin{array}{cc}\mp G_{\theta\theta} & \mp(G\mp B)_{\theta i} \\
                      0             &  1_{D-1}
\end{array}\right).
\ee
Using these, the transformation under duality of the metric and the
torsionful connections can be written as
\beq
\label{dmetric}
&\wt G^{-1} = Q_- G^{-1} Q_-^T = Q_+ G^{-1} Q_+^T ,& \\
\label{dconnec}
&\wt\Omega^{\pm M}_{NK}=(Q^{-1}_\mp)^{N'}_{~N}\,(Q^{-1}_\pm)^{K'}_{~K}
\,(Q_\mp)^M_{~M'}\, \Omega^{\pm M'}_{N'K'}
-\delta^i_K\,(\del_i\,Q_\mp\,Q^{-1}_\mp)^M_{~N}.&
\eeq
Now, consider the bosonic non-linear \s-model and let $p_\theta$
denote the momentum canonically conjugate to $\theta$. The duality
transformations (\ref{dual}) then follow from the canonical
transformation
\be
\label{ct-b}
\wt \theta'=-p_\theta\,, \qquad \wt p_\theta = - \theta'\,,
\qquad\wt x^i = x^i \,.
\ee
It is clear from the above that the relation between $\theta$ and
$\wt\theta$ is, in general, non-local.

In the following, we want to generalize the above procedure to the
case of $N=1$ supersymmetric non-linear \s-models defined by the
action:
\beq
\label{action}
S&=&{1\over 2}\int d^2 \sigma \big[ (G_{MN}+B_{MN})\del_+ X^M \del_-X^N
-i\psi_+^M G_{MN}(\delta_K^N\del_- + \Omega^{+N}_{LK}\del_-X^L)\psi_+^K
\nonumber \\
&-&i \psi_-^M G_{MN}(\delta_K^N\del_+ + \Omega^{-N}_{LK}\del_+X^L)\psi_-^K
+{1\over 2} \psi_+^M \psi_+^N \psi_-^K \psi_-^L R_{MNKL}(\Omega^-)
\big].
\eeq
Here, $R_{MNKL}(\Omega^\pm)$ are the curvature tensors corresponding
to the torsionful connections $\Omega^{\pm M}_{NK}$. The above action
has a default $(1,1)$ supersymmetry under which the fields transform as
\be
\label{N=1}
\delta_\mp X^M =\pm i \epsilon_\mp\psi_\pm^M ,\quad
\delta_\mp\psi_\pm^M =\pm\del_\pm X^M\epsilon_\mp ,\quad
\delta_\mp\psi_\mp^M =\mp i\psi_\mp^N\epsilon_\mp
\Omega^{\pm M}_{NK} \psi_\pm^K .
\ee
To preserve this $N=1$ supersymmetry under duality, we have to
supplement the canonical transformation (\ref{ct-b}) (where $p_\theta$
is now defined using the $N=1$ supersymmetric action) with the
appropriate transformations of the worldsheet fermions. In terms of
the bosonic and fermionic coordinates, the resulting canonical
transformation can now be written as
\be
\label{ct}
\wt\psi_\pm^M = Q^M_{\pm N}\psi_\pm^N ,\qquad
\del_\pm\wt\theta=Q^\theta_{\pm M}\,\del_\pm X^M +i\psi^j_\pm\,\del_j
Q^\theta_{\pm M}\,\psi^M_\pm .
\ee
Using the $N=1$ superfields $\Phi^M$, this takes the form
\be
\label{superfield}
D_\pm\,\wt\Phi^M = Q^M_{\pm N}(\Phi)\,D_\pm\,\Phi^N .
\ee
The conservation of the $\theta$-isometry current leads to
$\del_+\del_-\wt\theta=\del_-\del_+\wt\theta$. This implies that, in
spite of the non-local relation between $\theta$ and $\wt\theta$, on
shell, the dual coordinate is a local function of the worldsheet
coordinates $\sigma^\pm$.

Since the fermion couplings in (\ref{action}) are entirely determined
by the $N=1$ supersymmetry, the canonically transformed action has the
same form as the original action (\ref{action}) with the backgrounds
$G_{MN}, B_{MN}$ replaced by their dual counterparts as given by
(\ref{dual}). Now, comparing the two actions, we obtain a compact
expression for the transformation of the generalized curvature tensor:
\be
\label{dualcurv}
Q_{+M}^{M'}Q_{+N}^{N'}Q_{-K}^{K'}Q_{-L}^{L'}\wt R_{MNKL}(\wt\Omega^-)
= R_{MNKL}(\Omega^-)
-2 G^{-1}_{\theta\theta}\del_{[N}^{}Q^{\theta}_{+M]}
\del_{[L}^{}Q^{\theta}_{-K]} .
\ee
\section{T-Duality and Non-Local Extended Supersymmetry on the\\
  Worldsheet}
Let $J^M_N$ denote a complex structure on the target space
($J^2 =-1$) and define $\psi^{(J)M}_\pm=J^M_{\pm N}\psi^N_{\pm}$. The
invariance of the action (\ref{action}), under the replacement
$\psi^M_\pm\rightarrow \psi^{(J)M}_\pm$, requires that $J^TGJ=G$ and
$\nabla^{\pm} J =0$. Here, the covariant derivatives contain the
torsionful connections. If these relations hold, then the theory admits
a second set of supersymmetry transformations, which are obtained from
the $N=1$ transformations (\ref{N=1}) by the same replacement
$\psi^M_\pm\rightarrow \psi^{(J)M}_\pm$.

The effect of T-duality on the extended worldsheet supersymmetries can
be studied by requiring that this second set of supersymmetry
transformations of the original theory imply a similar set of
transformations for the dual theory. When $\del_{\theta}J$ is not
necessarily zero, we find that under duality the complex structures
$J_{\pm}(\theta, x^i)$  transform to non-local objects $\wt J_{\pm}$
given by
\be
\label{Jnlocal}
\wt J_{\pm}([\wt\theta, x^i],x^i)
 = Q_{\pm} J_\pm(\theta[\wt\theta, x^i],x^i) Q^{-1}_\pm .
\ee
Here, $\theta[\wt\theta, x^i]$ is the usual notation for the
functional dependence of $\theta$ on $\wt\theta$ and $x^i$ with the
explicit relation given by the second equation in (\ref{ct}).
Note that when $\del_{\theta} J =0$, this reduces to the known
transformation of $J$ as obtained in \cite{H1,KIR}. However, in
general, $\wt J_\pm$ has a non-local dependence on the coordinates of
the dual target space $\{\wt X^M\}=\{\wt\theta,\,x^i\}$. The condition
of the covariant constancy of the complex structure now gets modified
to
\beq
&\del_\theta \wt J^M_{\pm N}+\wt G^{-1}_{\wt\theta\wt\theta}\,
\left(\wt \Omega^{\pm M}_{\wt\theta L}\,\wt J^L_{\pm N}
-\wt J^M_{\pm L}\,\wt\Omega^{\pm L}_{\wt\theta N}\right) =0,&
\nonumber\\
&\wt\nabla^\pm_i\,\wt J^M_{\pm N} \pm (\wt G\pm \wt B)_{\wt\theta i}
\,\del_\theta \wt J^M_{\pm N} =0.&
\label{nlcov}
\eeq
Note that these equations contain derivatives of $\wt J_{\pm}$ with
respect to $\theta$ and not with respect to the natural coordinate on
the dual space, which is $\wt\theta$.

Even when the extended supersymmetry becomes non-local under duality,
the extended superconformal algebra remains unchanged. However, this
algebra is now realized in terms of non-local supercharges, and the
representation becomes non-local. Such non-local representations in a
class of conformal field theories were constructed in terms of
parafermions in \cite{Kounnas}.  There are several explicit examples
known in which a part of the extended supersymmetry becomes non-local
under duality \cite{Bakas,BS,BOK}.

\section{Implications for Target Space Supersymmetry}
A configuration of the bosonic background fields admits $N=1$
space-time supersymmetry provided the supersymmetric variations of the
gravitino ($\Psi_M$) and dilatino ($\lambda$) fields vanish. Let us
consider $\delta\Psi_M=0$, which leads to the Killing spinor equation
\be
\label{gravitino}
\delta \Psi_M=\del_M\eta +{1\over 4}\left(\omega^{AB}_M -
{1\over 2} H_M^{~~AB}\right) \gamma_{AB}\,\eta =0 .
\ee
Here, $\omega_M^{AB}$ is the spin connection and $A,B$ are tangent
space indices. When the target space supersymmetry is a consequence of
an extended supersymmetry on the worldsheet, then the complex
structure $J$ and the Killing spinor $\eta$ are related by
\cite{Strominger}
\be
\label{ts-ws}
J^M_{+N} = \bar\eta\,\gamma^M_{~N}\,\eta .
\ee
The Killing spinor condition then implies that $\nabla^+_M
J^K_{+N}=0$. Equation (\ref{ts-ws}) can be used to study the effect of
T-duality on space-time supersymmetry. In the following, we describe
the three cases that may arise:

\noindent {\it Case 1}\,: Here, $\del_\theta\eta =0$, which implies
$\del_\theta J_+ =0$. In this case, $\eta$ is invariant under duality
(up to a possible local Lorentz transformation). The Killing spinor
condition and hence the supersymmetry are preserved.

\noindent{\it Case 2}\,: If $\del_\theta J_\pm \ne 0$, then
$\del_\theta \eta\ne 0$. In this case, the extended worldsheet
supersymmetry is non-locally realized after duality. Equation
(\ref{ts-ws}) then implies that in the dual theory $\eta$ is replaced
by a non-local object $\wt\eta$, given by
\be
\label{deta}
\wt \eta ([\wt\theta, x],x) = \eta (\theta[\wt\theta,x],x) .
\ee
The Killing spinor condition is modified to
\beq
&\del_\theta\,\wt\eta+{1 \over 4}\wt G^{-1}_{\wt\theta\wt\theta}\,
\left(\,\wt\omega^{AB}_{\wt\theta} - {1\over 2}
\wt H_{\wt\theta}^{~~AB}\,\right)\gamma_{AB}\,\wt\eta =0,&
\nonumber\\
&\del_i\wt\eta +{1\over 4} \left(\,\wt\omega^{AB}_i -
{1\over 2}\wt H_i^{~~AB}\,\right) \gamma_{AB}\,\wt\eta
+\,(\wt G+\wt B)_{\wt\theta i}\,
\del_\theta\,\wt\eta = 0.&
\label{nlgravitino}
\eeq
This indicates that the target space supersymmetry is no longer
realized in the conventional way.

\noindent{\it Case 3}\,: The only other possibility is when $\eta$
depends on $\theta$ in such a way that the $\theta$-dependences on the
right-hand side of (\ref{ts-ws}) cancel out, giving rise to a
$\theta$-independent $J$. In this case, in the dual theory, the
extended worldsheet supersymmetry is locally realized while the
associated target space supersymmetry has a non-local realization.

The realization of supersymmetry in cases 2 and 3 is highly
non-conventional. Due to their non-local nature, these transformations
make sense only when the coordinates are restricted to the string
worldsheet, and not at a generic space time point. Since the
background fields are invariant under supersymmetry, the non-locality
does not show up as long as we are looking at the vacuum
configurations. However, the supersymmetry will be non-locally
realized on the spectrum of fluctuations around these backgrounds,
which are the relevant quantum fields for the low-energy theory.

In string theory, equivalence under T-duality is a consequence of
the existence of both momentum and winding modes associated with the
compact coordinate $\theta$. It is well known that under duality these
modes are interchanged: The conserved momentum $P_\theta$ and the
winding number $L_\theta$ associated with the compact coordinate
$\theta$ (with non-trivial $\pi_1$) are given by $P_\theta =
\int_0^{2\pi} d\sigma p_\theta$ and $L_\theta=\theta(\sigma=2\pi)-
\theta(\sigma=0)$. Then, from the canonical transformation
(\ref{ct}), it follows that $\wt P_\theta=-L_\theta$ and $\wt
L_\theta=-P_\theta$.  Since the momentum and winding modes are
associated with the worldsheet coordinates $\tau$ and $\sigma$,
respectively, their interchange under duality is the origin of the
non-local relationship between $\theta$ and $\wt\theta$. This can be
easily seen when the backgrounds are flat and one can write $\theta=
\theta_L + \theta_R$, whereas $\wt\theta = \theta_L-\theta_R = \int
d\sigma^+ \del_+\theta - \int d\sigma_- \del_-\theta$.  As for the
behaviour of supersymmetry, note that the parameter $\eta(\theta,x)$
is sensitive to the string momentum and winding modes associated with
$\theta$. The non-locality in the dual theory arises from the fact
that the momentum and winding modes of the dual string enter
$\wt\eta$ not through $\wt\theta$ (which would have resulted in a
local spinor $\wt\eta (\wt\theta)$), but through the original
coordinate $\theta$.


\begin{thebibliography}{50}
\bibitem{K}
E. Kiritsis, {\it Nucl. Phys.} {\bf B405} (1993) 109;
E. \'{A}lvarez, L. \'{A}lvarez-Gaum\'{e} and Y. Lozano,
{\it Nucl. Phys.} {\bf B415} (1994) 71.
\bibitem{GRV}
A. Giveon, E. Rabinovici and G. Veneziano, {\it Nucl. Phys.}
{\bf B322} (1989) 167;
E. \'{A}lvarez, L. \'{A}lvarez-Gaum\'{e} and Y. Lozano,
{\it Phys. Lett.} {\bf B336} (1994) 183.
\bibitem{H1}
S. F. Hassan, {\it Nucl. Phys.} {\bf B454} (1995) 86.
\bibitem{KIR}
I. T. Ivanov, B. Kim and M. Ro\v{c}ek, {\it Phys. Lett.} {\bf B343}
(1995) 133; B. Kim, {\it Phys. Lett.} {\bf B335} (1994) 51.
\bibitem{BOK}
E. Bergshoeff, R. Kallosh and T. Ortin, {\it Phys. Rev.} {\bf D51}
(1995) 3009.
\bibitem{H2}
S. F. Hassan, {\it T-Duality and Non-local Supersymmetries},
  CERN-TH/95-98, hep-th/9504148.
\bibitem{BS}
I. Bakas and K. Sfetsos, {\it Phys. Lett.} {\bf B349} (1995) 448.
\bibitem{Bakas}
I. Bakas, {\it Phys. Lett.} {\bf B343} (1995) 103.
\bibitem{Kounnas}
C. Kounnas, {\it Phys. Lett.} {\bf B321} (1994) 26;
I. Antoniadis, S. Ferrara and C. Kounnas, {\it Nucl. Phys.} {\bf
  B421} (1994) 343.
\bibitem{Strominger}
A. Strominger, {\it Nucl. Phys.} {\bf B274} (1986) 253.
\end{thebibliography}
\end{document}